\documentclass[12pt]{article}
\usepackage{epsfig}
\def\A{{\cal{A}}}
\def\CS{{\cal{S}}}

\def\N{{\cal{N}}}
\def\F{{\cal{F}}}
\def\CO{{\cal{O}}}
\def\R{{\cal{R}}}
\def\Cf{f}
\def\y{{\cal{Y}}}

\def\<{{\langle}}
\def\>{{\rangle}}
\def\nn{\nonumber}
\def\es{{=}}
\def\al{{\alpha}}
\def\dal{{\dot\alpha}}
\def\tr{{\hbox{Tr\,}}}
\newcommand{\refs}[1]{(\ref{#1})}

\def\Res{{\hbox{Res}}}
\def\half{{{1\over 2}}}
\begin{document}
\begin{flushright}
LMU-ASC 25/05\\
\end{flushright}

\begin{center}
{\large\bf Low Energy Effective Action in N=2 Yang-Mills as an
Integrated Anomaly}

\vskip 0.5truecm

{\large{M. Magro}${}^1$ and {I. Sachs}${}^2$}\\
\vskip 0.5cm
${}^1$ Laboratoire de Physique,\\
\'Ecole normale sup\'erieure de Lyon, \\
46 All\'ee d'Italie, 69364 Lyon Cedex 07, France\\
Marc.Magro@ens-lyon.fr\\

\vskip 0.2cm

${}^2$ Arnold Sommerfeld Center, Department f\"ur Physik\\
 Ludwig-Maximilians Universit\"at, \\
Theresienstrasse 37, D-80333, M\"unchen, Germany\\
ivo@theorie.physik.uni-muenchen.de
\end{center}

\begin{abstract}
\noindent Based on chiral ring relations and anomalies, as
described by Cachazo, Douglas, Seiberg and Witten, we argue that
the holomorphic effective action in N=2 Yang-Mills theory can be
understood as an integrated U(1) anomaly from a purely field
theory point of view. In particular, we show that the periods of
the Riemann surface arising from the generalized Konishi anomaly
can be given a physical interpretation without referring to
special geometry. We also discuss consequences for the
multi-instanton calculus in N=2 Yang-Mills theory.
\end{abstract}

\section{Introduction}
Many of the exact results for effective actions in quantum field
theory can be understood as integrated anomalous Ward-identities
which are protected from higher order corrections. Well known
examples are the integration of the chiral (Weyl) anomaly for
massless fermions in two dimensions coupled to a gauge
(graviational) field or the Veneziano-Yankielowicz superpotential
in $\N \es 1$ Yang-Mills theory \cite{VY1}.  The integration
constant in turn reflects the choice of regularization and
dynamical scale. In the latter example this constant is fixed by a
one-instanton calculation \cite{1Inst,Pouliot}.  The purpose of
the present paper is to understand the exact results for the low
energy effective action of $\N \es 2$ Yang-Mills theory
\cite{9407087} as an integrated anomaly equation.


That the low  energy effective action for $\N\es 2$ Yang-Mills
theory can be obtained by integrating  the superconformal anomaly
equation \cite{West,MSW} has been shown a while ago using a rather
intricate series of arguments for gauge group $G=SU(2)$ for pure
gauge theory in \cite{9610026} and including fundamental matter in
\cite{FMRS2}. For $SU(2)$ without matter this can be understood as
follows: The superconformal anomaly equation together with the
$SL(2,{\bf Z})$ structure of the mass formula implies an ordinary
second order differential equation for the first derivative of the
prepotential $\F(\A,\Lambda)$ for the massless $\N \es 2$ vector
multiplet $\A$. What is more is that the parameters of this
equation are completely fixed by the weak coupling asymptotics.
The integration of this equation then determines the prepotential
  $\F(\A,\Lambda)$ uniquely. For higher rank gauge groups,
however, the superconformal anomaly equation of $\N \es 2$
Yang-Mills is not sufficient simply because it only provides a
single equation for a rank $G$ number of low energy fields.

On the other hand, recent work pioneered by Vafa and collaborators
  \cite{0103067,0103067b,DV} using results on geometric
transitions in string theory has lead, among other results, to a
new string theoretic derivation of the low energy effective action
in $\N \es 2$ Yang-Mills theory. The starting point is a certain
class of $U(N)$, $\N \es 1$ theories obtained from $\N \es 2$
Yang-Mills by adding a superpotential $W(\Phi)$ for the chiral
multiplet that breaks the gauge symmetry $U(N)\to
U(N_1)\times\cdots\times U(N_n)$.  This theory can be
geometrically engineered via D-branes partially wrapped over
certain cycles of a Calabi-Yau geometry. At low energies, the
effective theory has a dual formulation
 where the branes are replaced by fluxes. Furthermore, the dual
 theory is described
 in terms of the gluino condensate $S_i$ which, together with
 the massless
$U(1)$ vector multiplets $w^i_\al$ in the $U(N_i)$, form
an $\N \es 2$ vector multiplet
on the dual Calabi-Yau geometry \cite{0103067,0103067b}. %
The holomorphic part of the effective action for this multiplet is
that of an $U(1)^n$, $\N \es 2$ Yang-Mills theory spontaneously
broken to $\N\es 1$ with a superpotential $W_{eff}(S_i)$.
Furthermore, this effective superpotential for the glueball
superfield can be written as an integral of the holomorphic
$3$-form over a blown up $S^3$ in the Calabi-Yau geometry, or
equivalently, a period integral over a dual cycle in a genus
$g$-Riemann surface $\Sigma$. In this formulation, the effective
couplings of the $U(1)$ vector multiplets appear directly as the
period matrix of this Riemann surface. Upon scaling the
 classical  superpotential $W(\Phi)$ to zero one then recovers the $\N\es 2$
theory at a point in the moduli space given by the minimum of
$W(\Phi)$. Although the $S_i$ vanish in the $\N\es 2$ limit the
structure of the Riemann surface $\Sigma$ survives and the limit
for the coupling $\tau_{ij}$ for the massless $U(1)$'s is smooth.

In this paper we investigate how this structure can be seen to
arise directly in the field theory formulation by integrating a
suitable anomalous Ward-identity. This program has in fact been to
a large extent completed in \cite{0211170} where a certain
generalization of the Konishi anomaly was used to establish that
the aforementioned Riemann surface $\Sigma$ arises directly in the
field theory limit. We should note that the prepotential for this
theory has been understood a while ago  via embedding in string
theory  \cite{kacru,9609239,9703166}  and more recently through
explicit multi-instanton computations to all orders
\cite{0206161,Flume}. On the other hand, the observation that the
effective action can be understood as an integrated anomaly may
put this result in its proper place within the exact results for
effective actions and also removes some of the mystery surrounding
this model. As we will explain, it also implies that the
multi-instanton calculus in this model is, in fact, equivalent to
an anomalous Ward-Identity together with a one-instanton
calculation and thus should elucidate the structure of the
multi-instanton calculus.

 The plan of this article is the following. In section 2 we
review the chiral ring relations of
 \cite{0211170} relevant for our program and discuss the
 consequences of these relations for instanton calculus
 in the $\N=2$ theory.  In particular we argue in section 2.3
 that a specific one-instanton calculation
 in the chiral ring together with an anomalous Ward-Identity
 completely determines the $n$-instanton contributions to the
 various quantities relevant in the $\N=2$ theory.

 Then, in section 3 we recall first the  constructions of the
prepotential $F(S_i,g_k)$ and the effective couplings $t_{ij}$ for
the massless $U(1)$ vector multiplets in terms of the Hessian of
$F(S_i,g_k)$ w.r.t.  $S_i$.
The key difference with the effective prepotential
$\F(\A,\Lambda)$ for the $\N\es 2$ theory on the Coulomb branch is
that while $\F(\A,\Lambda)$ has an infinite expansion in $\Lambda$
and thus involves {\em a priori} an infinite number of instanton
contributions, $F(S_i,g_k)$ is a homogenous function of the
"glueball" fields $S_i$ and the couplings $g_k$ of the
superpotential. In particular, the role of the dynamical scale
$\Lambda$ is reduced to setting the scale for the microscopic
gauge coupling  which can be evaluated in perturbation theory and
a one-instanton calculation.

What remains to be shown is that the Hessian of $F(S_i,g_k)$ is
given by the period matrix of the Riemann surface $\Sigma$. We
should note that while in the string theory description via
geometric transition \cite{0103067,0103067b} this is an immediate
consequence of special geometry, the proof in the field theory
description is more intricate. In sections  3.2-3.4 we will show
explicitly that this relation holds in the field theory
description. This is the main technical result of this paper.  For
$U(N)$ broken to $U(1)^N$ and in the limit of vanishing
superpotential, it allows us finally to express the low energy
effective couplings for the massless $U(1)$ vector multiplets in
terms of the $\N \es 2$, $U(N)$ Casimirs $u_k$ and in this way we
derive the low energy effective action \cite{9407087} for $\N\es
2$ Yang-Mills theory. This is done in section 3.5. Hence, the low
energy effective action has been shown to follow from integrating
an anomalous Ward-identity.

%


\section{Chiral Ring Relations}\label{SCR}

Our starting point is $\N \es 2$ $U(N)$ Yang-Mills  broken to an
$\N \es 1$,  $U(1)^N$  theory. In $\N \es 1$
superspace\footnote{We use the conventions of Wess and Bagger
\cite{Bagger}} this theory is described by a chiral multiplet
$\Phi$ and a vector multiplet $W_\alpha$. We add to the
corresponding action a classical superpotential $\tr W(\Phi)$ for
the chiral multiplet with
\begin{equation}
W(\Phi) = \sum_{k=0}^n {g_k \over k+1} \Phi^{k+1}. \label{wphi}
\end{equation}
As explained in \cite{0211170} an important role in the field
theory derivation of the glueball potential is played by the ring
of $\N \es 1$ gauge-invariant chiral operators. Since we will
rely on these properties as well we present in this section a
short summary of those that are relevant for us and discuss
some consequences.

\subsection{Definition}

The chiral ring is made up of gauge invariant operators $\CO$
which are chiral, ie.  $[Q_{\dal}, \CO\}=0$. It follows
immediately from this property that correlation functions of
operators in the chiral ring in a supersymmetric vacuum
($Q_\al|0\>=0$) are $x$-independent, ie.
\begin{equation}\label{one}
-2i\frac{\partial}{\partial
x_1^{\al\dal}}\<\CO_1(x_1)\CO_2(x_2)\>= \<[Q_\dal,[Q_\al,
\CO_1(x_1)\}\}\CO_2(x_2)\>=0\,.
\end{equation}
An important consequence of \refs{one} is the factorization
property $\< \CO_1 \CO_2 \> = \< \CO_1 \> \< \CO_2 \>$. Another
property of chiral operators is that in a supersymmetric vacuum
the expectation values of $\CO$ and $\CO+ Q_\dal X^\dal$
are identical provided $X^\dal$ is some gauge invariant operator.
Thus, we consider the chiral ring as being generated by the
equivalence classes
\begin{equation}
 \CO\simeq \CO+ Q_\dal X^\dal.
\end{equation}
One can represent (not necessarily gauge invariant) chiral
operators as the lowest components of chiral superfields
$\Psi$  (since $[Q_\dal,\Psi\}|_{\theta=0}=
[D_\dal,\Psi\}|_{\theta=0}$). Then, using
\begin{equation}
\bar \nabla^2\nabla_\al\Psi=[W_\al,\Psi\}\,,
\end{equation}
where $\nabla_\al$ is the super- and gauge covariant derivative,
one finds that $W_\al$ (anti) commutes with $\Psi$ in the
chiral ring. Taking this into account one sees that the $\N\es 1$
chiral ring is generated by the elements
\begin{eqnarray}\label{CR}
\{\tr(\Phi^k), \tr(\Phi^k W_\al),\tr(\Phi^kW_\al W^\al)\}\,.
\end{eqnarray}

In the rest of this paper certain relations between elements of
the chiral ring will play a crucial role. In order to describe
these relations for the complete set of elements simultaneously it
is useful to introduce a generating function for all elements of
the chiral ring. Such a generating function is given by
\begin{equation}
\R(z,\psi)\equiv\tr\hat \R(z,\psi)=
\frac{1}{2}\tr\left((\frac{1}{4\pi}W^\al-\psi^\al)^2
\frac{1}{z-\Phi}\right)\,, \label{eqlast}
\end{equation}
where $z\in {\bf C}$ and $\psi_\al$ is a Grassman-valued
parameter. The various elements of the chiral ring are then given
by the coefficients in the expansion of $\< \R(z,\psi) \>$ in
powers of $\psi_\al$ and $\frac{1}{z}$.

\subsection{Classical and Quantum Relations}
As explained in the introduction the basic idea we will employ is
to integrate an anomalous Ward identity in order to obtain the
effective action for the massless degrees of freedom in the
Coulomb branch of $\N\es 2$ Yang-Mills theory. Concretely this
anomaly is manifested as a quantum correction to a classical
relation in the chiral ring. The relation we consider is simply a
consequence of the equation of motion,  $\bar
\nabla^2\bar\Phi=\partial_\Phi W(\Phi)$,
where $W(\Phi)$ is the classical superpotential. Since classically
$\bar \nabla_\dal$ commutes with any chiral field we can multiply
this equation by $\hat\R$ to get
\begin{equation}
\bar D^2\tr\left(\hat \R(z,\psi)\bar\Phi\right)=
\tr\left(\hat \R(z,\psi)\partial_\Phi W(\Phi)\right)\,.
\end{equation}
Upon quantization, normal ordering effects have to be taken into
account which lead to an anomaly in the above relation
\cite{Konishi1,Konishi2}. This anomaly can be determined, for
instance, using  Pauli-Villars regularisation  (e.g. \cite{1001}).
In perturbation theory this one loop calculation is exact. It was
shown in \cite{0311238} that for the types of superpotentials
considered here there are no non-perturbative corrections to the
anomaly. The anomalous relation is then given by \cite{0211170}
\begin{equation}\label{AR}
\bar D^2\tr\left(\hat \R(z,\psi)\bar\Phi\right)=
\tr\left(\hat \R(z,\psi)\partial_\Phi W(\Phi)\right)+
\frac{1}{32\pi^2}\sum_{kl}\left[W_\al,
\left[W^\al,\frac{\partial \hat \R}{\partial\Phi_{kl}}\right]
\right]_{lk}\,,
\end{equation}
where $\Phi_{kl}$ are the entries of the matrix $\Phi$. Note
that the above equation still holds if we replace $W_\al$ by
$W_\al-4\pi\psi_\al$ since  $\psi_\al$ (anti) commutes with
everything. Still following \cite{0211170} and using
the identity
\begin{equation}
\sum_{kl}\left[\chi_1,\left[\chi_2,\frac{\partial }
{\partial\Phi_{kl}}\frac{\chi_1\chi_2}{z-\Phi}\right]\right]_{lk}=
\tr\left(\frac{\chi_1\chi_2}{z-\Phi}\right)
\tr\left(\frac{\chi_1\chi_2}{z-\Phi}\right)\,,
\end{equation}
valid in the chiral ring for anti commuting operators  $\chi_1$ and $\chi_2$
and taking expectation values, we have
\begin{equation}
\<\R(z,\psi)\R(z,\psi)\>=\tr\<\hat \R(z,\psi)\partial_\Phi
W(\Phi)\>.
\end{equation}
Finally, recalling the factorization properties of the chiral ring
this leads to \begin{equation}\label{RQ}
 \<\R(z,\psi)\>^2=\partial_z W(z)\<\R(z,\psi)\>+
 \frac{1}{4}\Cf(z,\psi)\,
\end{equation}
where $\Cf(z,\psi) = \sum_{i=0}^{n-1} f_i(\psi) z^i$ is a
polynomial of order $n-1$ in $z$ defined by
\begin{equation}
\Cf(z,\psi) =-2 \<\tr\left(\frac{(W'(z)-W'(\Phi)) ({W^\al \over 4
\pi} -\psi^\al)^2}{z-\Phi}\right)\>.
\end{equation}
The coefficients $f_i(\psi)$ are thus given by
\begin{equation}
f_i=-{1 \over 2 \pi} \sum\limits_{q=i+1}^ng_qt_{q-i-1} \label{fi}
\end{equation}
where we have defined
\begin{equation}\label{tkdef}
t_k(\psi)=\frac{1}{4\pi}\<\tr\left(\phi^{k}({W^\al } -4\pi
\psi^\al)^2\right)\>\,.
\end{equation}
 Eqn. \refs{RQ} is then solved for $\<\R(z,\psi)\>$ as
\begin{equation}\label{RL}
2\<\R(z,\psi)\>= W'(z)-\sqrt{W'(z)^2+\Cf(z,\psi)}\,.
\end{equation}
In the above expectation values it is understood that the high
energy degrees of freedom have been integrated out while the low
energy degrees of freedom play the role of background fields which
are assumed to take values compatible with unbroken supersymmetry,
so that the chiral ring properties are realized. It has been shown
in \cite{svrcek2} that each solution of the chiral ring relations
corresponds to a supersymmetric vacuum of the gauge theory. In
what follows we consider the case where the $U(N)$ gauge symmetry
is broken down to
 $U(1)^N$. In analogy with the case where the breaking is to a
 product of $U(N_i)$ groups, we take the low energy fields  to
 be the massless $U(1)_i$ vector multiplets $w_i^\al$ together
 with the glueball superfields\footnote{It is generally assumed
 that the glueball field is the appropriate variable in the low
 energy theory although we are not aware of a rigorous
 justification of this assumption within field theory
 (see also \cite{0311181}).} $S_i$ with the condensate of the
 massless gluinos as their lowest
 components. In particular, the chiral multiplet which acquires
 a mass due to the classical superpotential $W(\Phi)$ is
 integrated out. The light degrees of freedom are then
 conveniently combined  with the help of the auxiliary
 variable $\psi_\al$ as
  \begin{eqnarray}
\CS_i&=&S_i - w_i \psi + \half \psi^2
N_i \label{m1}\\
&=&\frac{1}{2\pi i}\oint\limits_{A_i}\<\R(z,\psi)\>\,,\nonumber
\end{eqnarray}
where the contour $A_i$ is around the cut extending from the
$i$-th minimum of the function $W(z)$ which are assumed to be
non-degenerate. The details concerning the projection onto $S_i$
in terms of the contour integrals can be found in \cite{0211170}.

It follows from \refs{m1} that the expectation
value $\<\R(z,\psi)\>$ depends on $\psi_\al$ only implicitly
through the $\CS_i$. Furthermore, defining
\begin{equation}\label{Reim1}
\y(z,\psi)=\sqrt{W'(z)^2+\Cf(z,\psi)}
\end{equation}
and using eq.\refs{RL} gives
\begin{equation}\label{Si}
\CS_i=-\frac{1}{4\pi i}\oint\limits_{A_i}\y\,.
\end{equation}
Note also that $\CS_i$ can be considered as a function of the variables $g_k$
and $f_j(\psi)$ and conversely $f(z,\psi) = f(z,\CS_j(\psi),g_k)$.

The result \refs{RL}, obtained in \cite{0211170} shows the
appearance of a genus $g=N-1$ Riemann surface $\Sigma$ defined by
eq.\refs{Reim1} (evaluated at $\psi_\alpha=0$) as a consequence of
the anomalous relation \refs{AR} in the chiral ring. The period
matrix $t_{ij}(S_i, g_k)$ of $\Sigma$ will be identified with the
low energy effective couplings $\tau_{ij}(u_k)$ of the massless
fields in the Coulomb branch of the $\N\es 2$ theory. In order to
establish this result we first need to understand how $t_{ij}(S_i,
g_k)$ enters in the effective action for the massless vector
multiplets $w_{i\al}$. This is done in section 3.


\subsection{Chiral Ring Relations and Instanton Calculus}
As explained above the expectation values in Eqn. \refs{RQ} stand
for integrating out the high energy degrees of freedom in the
presence of the "external" fields $S_i$ and $w_{i\al}$. However,
another interpretation is possible in which all degrees of freedom
are integrated out. In this case the coefficients $f_i$ defined by
eq.\refs{fi} are functions of $g_k$ only. If the superpotential is
such that the gauge symmetry is broken to $U(1)^N$, then there is
no strong infrared
dynamics  
such that the coefficients $f_i$ can be evaluated by treating the
superpotential perturbatively in the $\N=2$ theory. In addition
$f_i$, or rather $t_k=\frac{1}{4\pi}\tr\<\phi^{k}W^2\>$ (see Eqn.
\refs{tkdef}) is saturated by classical and instanton
contributions. This is because $t_k$ is annihilated by $\bar
Q_{\dal}$ implying  that the path integral localizes on just these
configurations.  Concretely we have for the $s$-instanton
contribution
\begin{equation}\label{fkeval}
t_k= \sum\limits_{r_0,\cdots,r_N} a_{s,r_0,r_1,\cdots,r_N}
\Lambda^{2Ns}g_0^{r_0}\cdots g_N^{r_N}\,,
\end{equation}
with $r_l\geq 0$ and subject to the constraints $\sum_l r_l=1$ and
$2Ns-\sum_l r_l(l+1)=k$.  These selection rules can be understood
from symmetries \cite{0211170} or alternatively by looking at the
details of the $s$-instanton calculation: Since we are expanding
about the theory with vanishing superpotential ($g_k=0$) only
non-negative powers of $g_k$ appear in the expansion\footnote{This
is in contrast to explicit instanton calculations in the Coulomb
branch of $\N=2$ Yang-Mills theory (see
\cite{Pouliot,Dorey,Yung,0206161,Flume} and references therein)
where gauge and dilatation symmetry is spontaneously broken from
the outset.}. In order to get a non-vanishing contribution from an
instanton with topological charge $s$ we need to saturate the
$4Ns$ fermionic zero modes:
    Two gluino zero modes appear in
$\tr\left(\phi^{k}\lambda^2\right)$. Further zero modes appear by
expanding in terms of the Yukawa coulings
$\tr(\lambda\phi^\dagger\psi)$. It is then not hard to see that in
order to produce    $2Ns$  gluino $\lambda$ and "quark" $\psi$
zero modes we need to have exactly one contribution
$g_l\tr(\phi^{l-1}\psi^2)$ from the potential $\int d^2\theta
W(\Phi)$. This explains the first selection rule given above.
Furthermore, the total number of scalars contained in $t_k$ and
$g_l  \tr(\phi^{l-1}\psi^2)$ must match the $2Ns-2$ scalars
coming from the expansion in the Yukawa couplings. This then leads
to the second selection rule stated above which now has the
interpretation of a tree level amplitude in the presence of $4Ns$
fermionic zero modes.  A consequence of these selection rules
is that the $t_k$'s vanish for $0 \leq k < N-1$ and $t_{N-1}$ is
determined by a 1-instanton contribution.


For $k\geq N$ the connected part of $t_k={1 \over 4 \pi}
\<\tr(W^2\Phi^k)\>$ is still subject to the selection rules  above
and similarly for the connected part of $u_k=\<\tr\Phi^k\>$, that
is
\begin{equation}\label{ukeval}
\<\tr\Phi^k\>_c= \sum_s a_{s}\Lambda^{2Ns}\,.
\end{equation}
Here the connected part of the expectation value is the part that
cannot be written as a product of expectation values with less
than $k$ fields. In particular, for $k<2N$, $u_k$ is fully
determined by its classical expression in terms of the $g_k$'s.

Combining the selection rules for \refs{fkeval} and \refs{ukeval}
with the chiral ring relation \refs{RL} one then concludes that
the expansions of $t_k$ and $u_k$ in the dynamical scale $\Lambda$
are  completely determined by a single $1$-instanton calculation
and the classical expressions for the $u_k$, $k\leq N$. Indeed,
expanding both sides of eq.\refs{RQ} in $1/z$ we obtain the
recurrence relation
\begin{equation}
\sum\limits_{p,q=0}^\infty \R_p\R_qz^{-p-q-2}=
\sum\limits_{q=0}^\infty\sum\limits_{p=0}^N g_p\R_qz^{-p-q-1}
\end{equation}
where  $\<\R(z,\psi)\> = \sum z^{-q} \R_q$.  The recursion
relation is then obtained by comparing equal powers in $z$ (see
also \cite{svrcek2} for a related discussion).

 An important application of this discussion is that, of the
coefficients $f_i$, only $f_0$ is non-vanishing \cite{0211170} and
is then equal to $ d_0 g_N^2\Lambda^{2N}$.  The factor $d_0$ can
in principle be determined by evaluating $t_{N-1}$ in the explicit
one-instanton computation.  Instead we will determine $d_0$ below
 by comparison with the $1$-instanton contribution to the
superconformal anomaly  in the $\N \es 2$ theory. This, with the
help of \refs{RL}, fixes $\<\R(z,\psi)\>$ and, in particular,
$S_i$ completely in terms of the $g_k$ and the one-instanton
coefficient $f_0$. This suggests that the chiral ring relations,
or equivalently the Konishi anomaly imply a relation between
instanton contributions of different topological charges (see
\cite{9506102,9805127,0405117} for other approaches to instanton
recursion relations).

\section{Effective Action and Low Energy Couplings}


In order to determine the couplings of the massless U(1) vector
multiplets $w_{\alpha}^i$, we will need to determine the
holomorphic part of the effective action for these fields. In the
$\N=2$ theory this action is usually expressed in terms of a
prepotential $\F(\A,\Lambda)$ for the $\N=2$ vector multiplet
$\A$. However, as explained in the introduction this prepotential
 involves an infinite number of instanton contributions. The key
observation \cite{DV} allowing to circumvent this problem is that
the low energy effective couplings can equally be obtained from an
effective superpotential $W_{eff}(g_k,S_i,w_{i\al})$  which in
contrast to $\F(\A,\Lambda)$ does not receive higher instantons
contributions.

\subsection{Effective Action}
In analogy with the usual effective
action in field theory the holomorphic part of the effective
action is given by the sum over 1PI graphs with $S_i$ and
$w_{i\al}$ insertions.
In fact we can say more. A "non-renormalisation theorem" given in
\cite{0211170} shows that in perturbation theory the only
contributions, compatible with the expected symmetries of the
effective action come from planar graphs. Furthermore these
planar graphs have either exactly one $w_{i\al}$ insertion at
two of the index loops and one $S_i$ insertion at each of the
remaining index loops or one $S_i$ insertion in all but one index
loop. The index loop without $S_i$ insertion being proportional
to $N_i$, where
\begin{equation}
N_i=\frac{1}{2\pi i}\oint\limits_{A_i}\<\tr\frac{1}{z-\Phi}\>
\end{equation}
counts the degeneracy of the vacuum corresponding to the $i-th$
minimum of the potential. This observation then implies that the
holomorphic part of the effective action for the low energy fields
can be expressed in terms of a single function $F(g_k,S_i)$.
Indeed let $F(g_k,S_i)$ be the sum over all 1PI graphs with
exactely one $S_i$ insertion at each of the index loops. Then,
since the effective potential is obtained  by either replacing one
$S_i$ by $N_i$ or two $S_i$ insertions by $w_{i\al}$'s, we have
\begin{equation}\label{FW}
W_{eff}(g_k,S_i,w_{i\al})=\sum_i N_i\frac{\partial
F(g_k,S_i)}{\partial S_i}+ \sum_{ij} \frac{\partial^2
F(g_k,S_i)}{\partial S_i\partial S_j}
 w_{i\al} w_j^\al \, .
\end{equation}
This latter relation can be written equivalently as
\begin{equation}\label{FWPSI}
W_{eff}(g_k,S_i,w_{i\alpha})=\F(g_k,\CS_i)\biggl\vert_{\psi^2}\,.
\end{equation}
where $\F(g_k,\CS_i)$ is obtained from $F(g_k,S_i)$ simply by the
substitution $S_i\to \CS_i$. In order to evaluate the coupling for
the massless $U(1)$ fields $w_{i\al}$  we then set the $S_i$ "on
shell".  There are two equivalent ways of implementing this. In
the last section we have seen that  on shell the $f_i$'s are
expressed in terms of $g_k$. On the other hand, from the point of
view of $W_{eff}(g_k,S_i,w_{i\alpha})$ this means that
\begin{equation}
\frac{\partial W_{eff}(g_k,S_i,w_{i\alpha}=0)}{\partial S_i} =0
\end{equation}
so that the coupling matrix is
\begin{equation}\label{mat}
t_{ij}(S_i, g_k)= \frac{\partial^2 F}{\partial S_i \partial S_j}.
\end{equation}

However, to compare \refs{mat} with the couplings for the massless
$U(1)$ vector multiplets in $SU(N)$ $\N\es 2$ Yang-Mills we need
to change variables from the $w_{\al i}$ to usual Cartan-Weyl
basis for $SU(N)$. We will call these variables $\al_i$,
$i=1,\cdots , N-1$. They are defined by
\begin{equation}
 \alpha_i =S_i - {1\over N} \sum_{j=1}^N S_j \quad \mbox{for}\quad
i=1,\cdots,N-1 \label{sal}
\end{equation}
while $\alpha_+ = {1 \over N} \sum_{i=1}^N S_i$.  In particular,
$\alpha_+ $ vanishes on shell. The couplings for the $\alpha_i$
are then  given by  $\tau_{ij}(g_k,S_i)={\partial^2 F \over
\partial \alpha_i \partial
\alpha_j}$.

 In principle $F(g_k,S_i)$ can now be computed directly order
by order in perturbation theory\footnote{As already emphasized in the introduction  $F(g_k,S_i)$ is a homogenous function in $g_k$ and $S_i$. In particular the instanton contributions are summarized in the $g_k$-independent integration constant which is fixed by a one-instanton calculation. See also \cite{0311181} for a discussion of the ambiguities in the field theory computation of $F(S_i,g_k)$.} \cite{0211017} (or in an associated matrix model
\cite{0206255}). However,  for the couplings,  we only need its
second derivative $F(g_k,S_i)$ with respect to  $S_i$.  We will
now show  explicitely that the Hessian of $F$ can be expressed in
terms of the period matrix of the Riemann surface described by
\refs{Reim1}. This is the main technical result of the present
paper.

\subsection{Effective Action and Chiral Ring}

First we recall  a result in \cite{0211170} relating the
derivative of $F(g_k,S_i)$ with respect to $g_k$ to the
expectation value of $\R(z,\psi)$ introduced in section \ref{SCR}.
To start with  we have from eq.\refs{wphi}
\begin{equation}\label{Wgk}
\frac{\partial W_{eff}(g_k,S_i,w_{i\alpha})}{\partial g_k}=
 \frac{1}{ (k+1)}\<\tr \left(\Phi^{k+1}
\right)\>_{\{S_i,w_{j\al}\}}\,.
\end{equation}
On the other hand $W_{eff}(g_k,S_i,w_{i\alpha})$ is expressed in
terms of $F(g_k,S_i)$ via eq.\refs{FWPSI}.  The claim is that this
implies
\begin{equation}\label{Peffphi}
\frac{\partial\F(g_k,\CS_i)}{\partial g_k}= -\frac{1}{2(k+1)}
\<\tr \left(\Phi^{k+1} \left({W_\al \over 4\pi} -  \psi_\al\right)
\left({W^\al \over 4 \pi} -  \psi^\al\right)
\right)\>_{\{S_i,w_{j\al}\}}\,.
\end{equation}
The identity \refs{Peffphi} follows from the observation that the
right hand side
depends on  the background fields $\{S_i,w_{j\al}\}$ only in the
combination $\CS_i$. On the other hand \refs{FWPSI} and \refs{Wgk}
imply that the $\psi^2$ component of both sides of the equation
\refs{Peffphi} agree. Finally, by construction both sides vanish
at $\CS_i=0$. Consequently the two functions are the same.

We will now show that $\frac{\partial
 F(g_k,S_i)}{\partial \alpha_i}$ is expressed
 in terms of the (dual) $B_i$-periods of the Riemann
surface $\Sigma$ defined by eq. \refs{Reim1}. In the string theory
description of this model \cite{0103067b} this property follows
directly from the special geometry of CY spaces. In field theory
we are not aware of a previous derivation of this property (see
also comment at the end of section 4 in \cite{0211170}). The
strategy we use is to first compute $\frac{\partial^2
F(g_k,S_i)}{\partial {S_i}\partial {g_k}}$ and then to show that
the result obtained can be integrated to obtain $\frac{\partial
F(g_k,S_i)}{\partial {S_i}}$ up to an integration "constant"
independent of $g_k$.
 We start with eq. \refs{Peffphi} which gives for $\psi_\alpha =0$
\begin{eqnarray}
\frac{\partial F(g_k,S_i)}{\partial g_k}&=&-\frac{1}{32
\pi^2(k+1)}
\<\tr \left(\Phi^{k+1} W_\al W^\al\right)\>_{S_i}\nn\\
&=& -{1\over 2} {1\over (k+1)} Res_{\infty} \left[ z^{k+1} y(z)
\right]\,, \label{Fresy}
\end{eqnarray}
where we have used eq.\refs{eqlast}, \refs{RL} and
$y(z)={\cal{Y}}(z)|_{\psi_\al=0} = \sqrt{W'(z)^2 + f(z)}$. The
next step is then to characterize completely $y(z)$. As it is a
mereomorphic $1$-form with a singular point of order $N$ at
infinity, it can be written as \cite{Russie}
\begin{equation}
y=\sum_{k=0}^N\beta_k d\Omega_k +\beta_{+} d\Omega_{+}+
\sum\limits_{i=1}^{N-1}\gamma_i\xi_i
\end{equation}
with the following definitions: $\xi_i$ are canonically normalized
holomorphic $1$-forms such that $\oint_{A_i}\xi_j=\delta_{ij}$.
$d\Omega_k$ behave at infinity like $d\Omega_k=z^k+O(z^{-2})$ and
satisfy\footnote{Note that this can always be achieved by adding a
suitable linear combination of holomorphic $1$-forms.}
$\oint_{A_i} d\Omega_k=0$
 for all $i$. We have similarly $d\Omega_{+}=z^{-1} +
O(z^{-2})$ and $\oint_{A_i} d\Omega_+=0$  for $i=1,\cdots,N-1$.

Let us now determine the coefficients $\beta_k$, $\beta_+$ and
$\gamma_i$. The coefficient $\gamma_i$ is obtained by integrating
around the contour  $A_i$. It follows from \refs{Si}, the previous
definitions and from \refs{sal} that $\gamma_i=-4\pi i
(\al_i+\al_+)$. The coefficient $\beta_k$ is obtained by
considering the behavior of $y$ at infinity. More precisely, as
\begin{equation}
y(z)=W'(z)\sqrt{1+\frac{f(z)}{W'^2(z)}}=W'(z)(1+O(z^{-N-1})),
\end{equation}
we have $\Res_{\infty}(z^{-k-1}y)=g_k$ for $0\leq k\leq N$ which
leads to $\beta_k=g_k$. Finally, we have $\beta_{+} = -2N
\alpha_+$ as
\begin{equation}
\beta_{+} = {1 \over 2 \pi i}\oint_{\infty}y=  {1 \over 2 \pi i}
\sum_{i=1}^N \oint_{A_i} y = -2 \sum_{i=1}^N S_i =  -2 N\alpha_+.
\end{equation}
Thus we have shown that \label{sectiony}
\begin{equation}
y = \sum_{k=0}^N g_k d\Omega_k - 4\pi i
\sum\limits_{i=1}^{N-1}\alpha_i \xi_i -  \alpha_+ (2N \, d\Omega_+
+ 4\pi i\sum\limits_{i=1}^{N-1} \xi_i). \label{whatisy}
\end{equation}
We prove then in appendix  $A$ that $y$ is homogenous of degree
one in $\alpha_i$, $\alpha_+$ (or $S_i$) and $g_k$. In particular,
we have
\begin{equation}
{\partial y \over \partial g_k} = d\Omega_k \quad \hbox{and} \quad
{\partial y \over \partial \alpha_i} = - 4 \pi i \xi_i.
\label{deriofy}
\end{equation}

After these preparations we consider  $\frac{\partial^2
F}{\partial
 {g_k} \partial {\alpha_i}}$. We get then from  equations
\refs{Fresy} and \refs{deriofy}
\begin{eqnarray}
\frac{\partial^2F}{\partial {g_k}\partial
{\alpha_i}}&=&\frac{\partial^2 F}{\partial {\alpha_i}
\partial{g_k}}= -{1
\over 2} \frac{1}{k+1}Res_\infty(z^{k+1}\frac{\partial y}
{\partial\alpha_i} ),\nn\\
&=& 2 \pi i \frac{1}{k+1}Res_\infty(z^{k+1}\xi_i) \,.\label{eq1}
\end{eqnarray}

In the next subsection we integrate this relation.

\subsection{Identification of the $B_i$ periods}

\begin{figure}[htbp]
\begin{center}
\centering
\includegraphics[height=3cm]{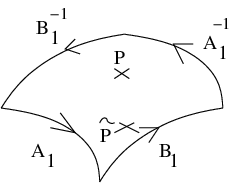}
\caption{ \label{figrblr}}
\end{center}
\end{figure}

 We will now show that the
expression \refs{eq1} can be written as a derivative with respect
to $g_k$ of an integral over the   cycle $B_i$ dual to $A_i$. For
this, we use the
 Riemann bilinear relations associated with one form of first
 kind (i.e. holomorphic), $\xi_i$, and the other of second kind
 (i.e.
 meromorphic with no residues), $d\Omega_k$:
 \begin{equation}
\sum_{j=1}^{N-1} \left[ \oint_{A_j} d\Omega_k \oint_{B_j} \xi_i -
\oint_{A_j} \xi_i \oint_{B_j} d\Omega_k \right]= 2 \pi i \left[
\Res_P(\Omega_k \xi_i) + \Res_{\tilde{P}}(\Omega_k \xi_i) \right].
 \label{rb}
 \end{equation}
Let us recall briefly that such relations are obtained by
computing in two different ways one particular integral
\cite{RBLR}, $\int_{\partial \Sigma } \Omega_k \xi_i$, where the
Riemann surface $\Sigma$ is thought here as a polygon with some
identifications (see figure \ref{figrblr}). Then, the l.h.s.
corresponds to the computation of this integral on the contour
shown on figure \ref{figrblr} while the r.h.s. of eq.\refs{rb} is
simply obtained by use of Cauchy's formula where $P$ and
$\tilde{P}$ have coordinates $z(P)=z(\tilde{P})=\infty$ with
$y(P)=-y(\tilde{P})$.

However,  for the l.h.s. we use the definitions of $\xi_i$ and
$d\Omega_k$, which give respectively $\oint_{A_j} \xi_i =
\delta_{ij}$ and $\oint_{A_j} d\Omega_k =0$. For the r.h.s., we
have $\Res_{P}(\Omega_k \xi_i)= \Res_{\tilde{P}}(\Omega_k \xi_i)$
because both $d\Omega_k ={\partial y \over \partial g_k}$ and
$\xi_i$ change sign when going from $P$ to $\tilde{P}$.
Furthermore these residues are equal to ${1 \over k+1}
\Res_\infty(z^{k+1} \xi_i)$ as $d\Omega_k = z^k + O(z^{-2})$ and
$\xi_i = O(z^{-2})$. Using the result \refs{deriofy} we finally
get
\begin{equation}
- {1 \over 2} \oint_{B_i}\frac{\partial y}{\partial g_k} = 2 \pi i
\frac{1}{k+1}\Res_\infty(z^{k+1}\xi_i).
\end{equation}
This enables us to write eqn.\refs{eq1} as $\frac{\partial^2 F}{
\partial{g_k} \partial{\alpha_i} }=-{1 \over 2}
\oint_{B_i}\frac{\partial
y}{\partial g_k}$ which can be integrated w.r.t. $g_k$ to end up
with
\begin{eqnarray}
\frac{\partial F}{\partial {\alpha_i}}=-{1 \over 2} \oint_{B_i}y+
H_i(\alpha_+,\alpha_j) \label{FBHi}
\end{eqnarray}
where $H_i(\alpha_+,\alpha_j)$ is an integration constant. The
purpose of the next subsection is to fix this function {\em via}
the superconformal anomaly.

\subsection{Determination of the Integration Constant}

In order to determine $H_i(\alpha_+,\alpha_j)$ we need to know the
$g_k$-independent part of $F(g_k,\al_i,\al_+)$. This can be done
with the help of the superconformal anomaly which can be computed
in the microscopic $U(N)$ theory using standard methods such as
Pauli-Villars regularization. This leads to
\begin{equation}
D^{\dal}T_{\al\dal}=-(3N_c-N_aN_c)D_{\al}\hat S\,,
\end{equation}
where $T_{\al\dal}$ is the $\N\es 1$ supercurrent and $\hat S$ is
the $SU(N)$ glueball field. More precisely, we have
\begin{equation}
S_i = \hat{S}_i - {1 \over 2 N} w_{\alpha i}w^{\alpha i}.
\end{equation}
This anomaly ought to be reproduced by the low energy effective
potential $W_{eff}(g_k,\hat S_i)$.  The charges of $\hat{S}_i,g_k$
and $\theta_\al$ are $3$, $2-k $, and  $-{1 \over 2} $
respectively. We thus conclude that $W_{eff}(g_k,\hat{S}_i)$
satisfies the equation
\begin{equation}
\left(\sum_k(2-k)g_k{\partial\over\partial
 g_k}+3\sum_i\hat S_i {\partial\over\partial \hat S_i}
 -3\right)W_{eff}=-2N\sum_i\hat S_i\label{w3}
\end{equation}
where we have assumed $N_a=1$  and $N\equiv N_c$ in the last
equality\footnote{ We note in passing
that the superconformal anomaly equation can be combined with the
Konishi anomaly $-\sum_k (k+1)g_k{\partial\over\partial g_k}
W=-2N\hat S$ to show that
\begin{equation}
\left(\sum_k g_k{\partial\over\partial
 g_k}+\sum_i\hat S_i {\partial\over\partial \hat S_i}
 -1 \right){W}=0\label{h3}
\end{equation}
i.e. $W_{eff}(g_k,S_i)$ is a homogenous function of degree $1$.}.
In order to see how this anomaly is reproduced in the low energy
effective theory we need to relate the dynamical scale $\Lambda_i$
of the low energy theory to the scale $\Lambda$ of the $U(N)$
theory \cite{0103067}. We have
\begin{eqnarray}\label{scales}
\Lambda_i^{3N_i}&=&
\Lambda^{2N}m_{\Phi_i}^{N_i}\prod\limits_{j\neq i}
m_{W_{ij}}^{-2N_j}\nn\\
&=&\Lambda^{2N}g_N^{N_i}\prod\limits_{j\neq i}(z_i-z_j)^{N_i-2N_j}
\,
\end{eqnarray}
  where $z_i$ are the roots of $W'(z)$. Taking into account the
dimensions of $g_k$ and $z_i$ it is then not hard to see that the
anomaly \refs{w3} is reproduced by
\begin{equation}
W_{eff}(g_k,\hat S_i)=\sum\limits_{i=1}^N
\hat{S}_i\log\left(\frac{\Lambda_i^{3N_i}}{\hat
S_i^{N_i}}\right)+P(g_k,\hat S_i) \label{eqqq}
\end{equation}
where $P(g_k,\hat S_i)$ is a  homogeneous polynomial in $g_k$ and
$\hat S_i$ transforming with weight $3$. On the other hand, from
\refs{FW} we have
\begin{eqnarray}
W_{eff}(g_k,\hat S_i)&=&\sum_iN_i\frac{\partial F}{\partial
{S_i}}\biggl\vert_{S_i=\hat S_i}. \label{apreqqq}
\end{eqnarray}
  Let $F_0(S_i)$ be the $g_k$-independent part of $F(S_i,g_k)$.
Then, eq.\refs{eqqq} and eq.\refs{apreqqq} imply that ${\partial
F_0 \over \partial S_i} = -S_i \log S_i+cS_i$ where $c$ is an
undetermined constant. Alternatively this gives  for $1 \leq i
\leq N-1$
\begin{equation}
{\partial F_0 \over \partial \alpha_i} = -S_i \log S_i +cS_i + S_N
\log S_N-cS_N \label{gkindependent}
\end{equation}
 where we have used the symmetry under the permutation of the
$S_i$.

\medskip

\begin{figure}[htbp]
\begin{center}
\centering
\includegraphics[height=3cm]{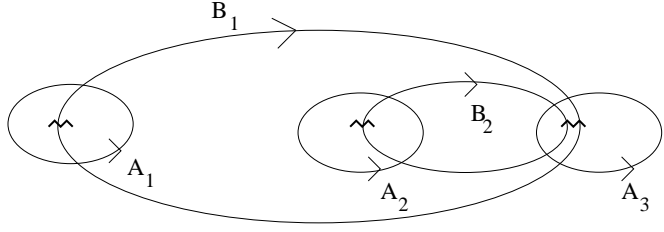}
\caption{Riemann Surface $\Sigma$ for $U(3)$ \label{mono}}
\end{center}
\end{figure}

Let us then compare this result with the $g_k$-independent term in
$-{1\over 2}\oint_{B_i}y$. This can be done using a scaling
argument (see also  \cite{0103067}). Suppose we rescale all
couplings $g_k$ by $\lambda$, so that $W'(z)\to \lambda W'(z)$,
$f(z)\to \lambda f(z)$ and consider $\tilde S_i=S_i/\lambda$ which
is thus given by
 \begin{eqnarray}\label{Sil}
\tilde S_i&=&-\frac{1}{2}\frac{1}{2\pi
i}\oint\limits_{A_i}\sqrt{W'^2+\frac{f}{\lambda}}\,.
\end{eqnarray}
If we then let $\lambda$ go to infinity,  $\tilde S_i$ goes to
zero. Geometrically this limit corresponds to vanishing $A_i$
cycles  since $ \frac{f}{\lambda}\to 0$. Therefore we can rotate
the two endpoints of the cut along one vanishing cycle $A_i$.
Under such a transformation, $S_i$ picks up a phase $e^{2\pi i}$.
 The transformation of the cycles can be worked out from
figure \ref{mono}:  under a monodromy transformation  $\tilde S_N
\rightarrow e^{2 \pi i} \tilde S_N$ all the $B_i$ change to $B_i
+A_N$ while under a monodromy transformation $\tilde S_i
\rightarrow e^{2\pi i} \tilde S_i$, with $1\leq i \leq N-1$,
only $B_i$ is changed in $B_i -A_i$.
This implies the asymptotic form of the period integrals
\begin{equation}
-{1\over 2} \oint_{B_i} \sqrt{W'^2+\frac{f}{\lambda}}\quad
{\stackrel{\lambda\rightarrow \infty}{=}} \quad-\tilde{S_i} \log
\tilde{S}_i + \tilde{S_N} \log \tilde{S}_N +
\tilde{P}_i(\tilde{S}_j,\tilde{S}_N)
\end{equation}
where $\tilde{P}_i(\tilde{S}_j,\tilde{S}_N)$ is an invariant
polynomial of $\tilde{S_j}$ and $\tilde{S_N}$ with no constant
term as the period integrals vanish in the limit
$\lambda\rightarrow \infty$.
Therefore, we get for large $\{g_k\}$ ($\lambda \to\infty$)
\begin{eqnarray}\label{in10}
-{1 \over 2} \oint_{B_i} y &=& \lambda \left( -{1\over 2}
\oint_{B_i} \sqrt{W'^2+\frac{f}{\lambda}} \right)\\
&_{\stackrel{=}{\lambda \rightarrow \infty}}& - S_i \log({S_i
\over \lambda}) + S_N \log({S_N \over \lambda}) + L_i(S_j,S_N) \nn
\end{eqnarray}
where $L_i(S_j,S_N)$ is a $g_k$-independent linear combination in
$S_j$ and $S_N$ since only the
 linear terms  of $\lambda \tilde{P}_i({S_j \over \lambda},
 {S_N \over \lambda} )$
 contribute in  the limit
$\lambda \rightarrow \infty$.
On the other hand we see from \refs{scales} that rescaling of
$g_k$ by $\lambda$ corresponds simply to rescaling of $\Lambda_i$
so that we get
\begin{eqnarray}\label{in110}
-{1 \over 2} \oint_{B_i} y  = - S_i \log({S_i \over \Lambda_i^3 })
+ S_N \log({S_N \over\Lambda_N^3 }) + \tilde{c}(S_i -S_N) + \cdots
\end{eqnarray}
where the dots refer to terms which depend on $g_k$ and where we
have used the property that any choice of the basis of cycles
$A_i$ is physically equivalent.

Thus, the comparison of eq.\refs{FBHi} on one hand with
eq.\refs{gkindependent} and eq.\refs{in110} on the other hand give
that the "integration constant" $H_i(\alpha_j,\alpha_+)$ is linear
or more precisely that
\begin{equation}
\frac{\partial F}{\partial {\alpha_i}}=-{1 \over 2} \oint_{B_i}y +
d_1 (S_i -S_N)\,. \label{eq53}
\end{equation}
The result \refs{eq53} enables us  then by use of
eq.\refs{deriofy} to obtain the low energy couplings
$\tau_{ij}(g_k)$ for the vector multiplets in terms of
  the period matrix of the Riemann surface $\Sigma$
defined by anomalous chiral ring relations \refs{Reim1}, that is
\begin{equation}\label{tauf}
\tau_{ij}(g_k)=\frac{\partial^2F(g_k,\al_i,\al_+)} {\partial
\al_i\partial \al_j}\biggl\vert_{\alpha_i=\bar \alpha_i}
 =2 \pi i \oint_{B_i} \xi_j +  d_1 \delta_{ij} + d_1\,,
\end{equation}
where $\bar \alpha_i$ denotes the value on shell. In particular
$\bar \alpha_+$ vanishes. Since the terms proportional to $d_1$ do
not match the structure of the $1$-loop correction, we have
$d_1=0$.
 Furthermore, using the result $f(z)=d_0 g_N^2 \Lambda^{2N}$ on
 shell, $\Sigma$ is given by
\begin{equation}\label{tauf2}
y^2(z)=W'^2(z)+d_0g_N^2\Lambda^{2N}\,.
\end{equation}
The constant  $d_0$  will be fixed in the next subsection.

\subsection{$\N \es 2$ Limit and Moduli of $\Sigma$}

To complete our derivation of the low energy couplings for the
massless U(1) vector multiplets $w_{\alpha}^i$ we  first need to
express the period matrix  in eq.\refs{tauf} in terms of the
Casimir variables $u_k=\<\tr \phi^k\>$, $(1\leq k \leq N$), of the
$\N \es 2$ theory \cite{9407087} and then fix the  constant $d_0$
appearing in the equation for the Riemann surface $\Sigma$.

 For this purpose, if $\phi_c$ is  the classical value of $\phi$
 obtained by extremizing the tree-level superpotential, we write
 $W'(z)=g_NP_N(z)$, where the coefficients of the polynomial
 $P_N(z)$ depend on $\tr (\phi_c^k)$ through the relation
\begin{equation}
P_N(z)=\det(z Id -\phi_c)\,.
\end{equation}
We know however from section 2.3. that there are no quantum
corrections to the Casimir variables i.e. that $u_k =
\tr(\phi_c^k)$. Thus, $P_N(z)=\<\det (z Id-\phi)\>$ and the
 equation \refs{tauf2} for $\Sigma$ becomes $y^2 =
g_N^2(P_N(z)^2 +d_0 \Lambda^{2N})$ or equivalently
\begin{equation}\label{y}
y^2= \<\det (z Id-\phi)\>^2 + d_0 \Lambda^{2N}.
\end{equation}
 The low energy couplings $\tau_{ij}(u_k)$ are thus given by the
period matrix of the Riemann surface $\Sigma$ defined by
eq.\refs{y}. Furthermore, since the expectation values of the
massless scalars are given by
\begin{equation}
a_i=\frac{1}{2\pi i} \oint\limits_{A_i}\<\tr ( {z \over z -
\phi})\>,
\end{equation}
we get from eq.\refs{eqlast} together with eq.\refs{y}
\begin{equation} \label{a}
a_i = {1 \over 2 \pi i} \oint\limits_{A_i} {z P^{\,'}_N(z) \over
y(z)} \,.
\end{equation}
The two equations \refs{tauf} and \refs{a} then determine the
prepotential $\F(a_i,\Lambda)$   via
$\tau_{ij}(a_i)=\frac{\partial^2}{\partial a_i\partial
a_j}\F(a_i,\Lambda)$ up to physically irrelevant integration
constants \cite{Naculich}.

Finally let us fix the constant $d_0$.  This can be done in
several ways. Here we will fix $d_0$ by comparing the prediction
for the observable $u_2(a_i)$  with the explicit one-instanton
calculation. Indeed, we have on one hand from the superconformal Ward identity \cite{9506102,9510129,West,MSW}
\begin{equation}\label{57}
\frac{iN}{\pi}u_2=2\F-\sum\limits_{i=1}^{N-1}a_i\partial_{a_i}\F\,.
\end{equation}
Upon substitution of the asymptotic instanton expansion for
$\F(a_i)$ this then leads to \cite{9506102,9609104}
\begin{equation}
u_2=\sum_i\phi_i^2+2\sum\limits_{k=1}^\infty k\F_k(a_i)\left(-{d_0
 \over 4} \right)^k \Lambda^{2Nk}\,, \label{58}
\end{equation}
with \cite{Ito}
$\F_1(a_i)=\sum\limits_{i=1}^N \prod\limits_{j\neq
i}\frac{1}{(\phi_i-\phi_j)^2}\,$.
Here $\phi_i$ stands for the diagonal entries of $\<\phi\>$. Eqn
\refs{58} is then a direct consequence of \refs{tauf} and
\refs{tauf2}. On the other hand, comparing \refs{58} with the
explicit one-instanton calculation \cite{Pouliot,Dorey,FT,9609104}
implies then $d_0=-4$.
Thus we have shown that the low energy couplings are indeed given
by the period matrix of $\Sigma$ with $d_0=-4$.

\section{Conclusions}

In this article we have  completed the field theory proof of the
claim that the holomorphic effective action of $\N \es 2$ $SU(N)$
Yang-Mills action can be obtained by integrating a suitable
anomaly. In particular, we have field the gap that arises in field
theory (not present in string theory) due to the absence of
special geometry relation between the periods of the Riemann
surface $\Sigma$.

In addition we have pointed out that the anomalous chiral ring
relations have important consequences for the multi-instanton
calculus. Indeed, the instanton contributions to the expectation
values of chiral ring elements are completely fixed by the
$1$-instanton observable $\< \tr (\Phi^{N-1} W^2) \>$. This also
applies in a similar way to the expectation value
$a_i=\frac{1}{2\pi i} \oint\limits_{A_i}\<\tr ( {z \over z -
\phi})\>$. On the other hand, expanding the contour integral in
$\Lambda$ one reads off the contribution of the $n$-instanton
contribution to the function $a_i(g_k)$ (or equivalently $a_i(u_k)
$). This seems to suggest a recursive structure that would be
interesting to investigate within the multi-instanton calculus.

\vspace{2cm}

{\bf Acknowledgments: }\\
It is a pleasure to thank Andreas Brandhuber, Fran\c{c}ois Delduc,
Dominique L\"ange,  Steffen Metzger, Guiseppe Policastro, Christian R\"omelsberger,
Samson Shatashvili and especially Peter Mayr for many helpful
discussions. The work of I.S. is supported by RTN under contract
005104 ForcesUniverse, by DFG 1096, string theory and by SFB 375
of the Deutsche Forschungsgesellschaft. M.M. acknowledges support
from the european network EUCLID-HPRNC-CT-2002-00325.

\section*{Appendix}

In this appendix, we prove that $y(z,g_k,S_i)=
\sqrt{W'(z)^2+f(z)}$ is homogeneous of degree 1 in $S_i$ and
$g_k$. For that purpose, we compute the various derivatives of $y$
and compare the results obtained with the expression
\refs{whatisy} of $y$ found in section \ref{sectiony},
\begin{equation}
y = \sum_{k=0}^N g_k d\Omega_k - 4\pi i
\sum\limits_{i=1}^{N-1}\alpha_i \xi_i -  \alpha_+ (2N \, d\Omega_+
+ 4\pi i\sum\limits_{i=1}^{N-1} \xi_i). \label{newwhatisy}
\end{equation}

The general strategy in the computations that will be done is to
identify the holomorphic contributions. Let us recall at this
stage that the space of holomorphic forms on the Riemann surface
$\Sigma$ is $N-1$ dimensional and two special basis are
$\{\xi_i\}$ and $\{ {z^j \over 2y} \}$ with $i=1,\cdots,N-1$ and
$j=0,\cdots,N-2$. We change then the variables from $(g_k,S_i)$ to
$(g_k,f_i)$ where $f_i(S_j,g_k)$ is the coefficient of order $i$
of $f(z)$. Indeed, ${\partial y \over
\partial f_i} = {z^i \over 2 y}$ is holomorphic except for
$i=N-1$. However, it is easy to prove that
\begin{equation}
f_{N-1}(g_k,S_i)=- 4g_N \sum_{i=1}^N S_i \label{derfn1}
\end{equation}
by computing the residue of $y$ at infinity as it follows
 from the definition of $y$ and the result
\refs{newwhatisy} that it is respectively equal to ${f_{N-1} \over
2 g_N}$ and to $- 2\alpha_+$.

We first compute the derivative w.r.t. $S_i$. We have
successively:
\begin{eqnarray}
- \half \left({\partial y \over \partial S_i}\right)_{S_j,g_k}
\!\!\!\!\!\!\!\!\!\!\!(g_k,S_j)&=& - \half
\sum_{j=0}^{N-1}\left({\partial y \over
\partial f_j}\right)
\left( {\partial f_j \over \partial S_i} \right)\nn\\
&=& -\half \sum_{j=0}^{N-2}\left({\partial y \over \partial
f_j}\right) \left( {\partial f_j \over \partial S_i} \right)
-\half \left({\partial y \over \partial f_{N-1}}\right) \left(
{\partial f_{N-1} \over
\partial S_i} \right)\nn\\
&=& \sum_{j=0}^{N-2} \left( {-z^j \over 4y} {\partial f_j \over
\partial S_i} \right)  + {  g_N z^{N-1} \over y}. \label{r3}
\end{eqnarray}
The first contribution in the r.h.s. of \refs{r3} is a linear
combination of holomorphic terms  while the second term is a
meromorphic form that has residue $1$ at infinity.

Consider then the case $i=N$. In that case, we also have from the
definition \refs{Si} of $S_i$
$$
 \oint_{A_i} - {1\over 2}
\left({\partial y \over \partial S_N}\right)= 2 \pi i
\left({\partial S_i \over
\partial S_N}\right)_{S_i,g_k}
\!\!\!\!\!\!\!\!\!\!\! =0
$$
for $i=1,\cdots, N-1$. Thus $-{1\over 2} \left({\partial y \over
\partial S_N}\right)$ is a 1-form whose $A_i$ integrals for
$i=1,\cdots, N-1$ vanish and which has residue $1$ at infinity.
However, there is an unique form satisfying these properties and
by definition it is $d\Omega_{+}$. Thus, we have shown that
\begin{equation}
- {1 \over 2} \frac{\partial y}{\partial S_N}= d\Omega_{+}.
\end{equation}
Take then $1\leq i\leq N-1$. For $1\leq j \leq N -1$, we get
$\delta_{ij} = {1 \over 2 \pi i} \oint_{A_j} - \half ({\partial y
\over \partial S_i})$. Thus, if we define the $1$-form $k_i =
-\half {\partial y \over \partial S_i} - 2\pi i \xi_i$ we have on
one hand $\oint_{A_j} k_i =0$ for $1\leq j \leq N-1$ and on the
other hand $\Res_\infty k_i = \Res_\infty (-\half {\partial y
\over \partial S_i}) = 1$. Thus, for the same reason as above we
have  $k_i = d\Omega_+$ and so
\begin{equation}
-\half \frac{\partial y}{\partial S_i}= 2\pi i \xi_i + d\Omega_{+}
\quad \hbox{for} \quad 1\leq i\leq N-1.
\end{equation}
It is now easy to compute the derivatives w.r.t. $\alpha_i$ and
$\alpha_+$ to get
\begin{eqnarray}
{\partial y \over \partial \alpha_i} &=&  {\partial y \over
\partial
S_i} - {\partial y \over \partial S_N}= - 4 \pi i \xi_i,
\label{derai}\\
{\partial y \over \partial \alpha_+} &=& \sum_{i=1}^{N} {\partial
y \over \partial S_i}= - 4 \pi i \sum_{i=1}^{N-1} \xi_i - 2 N \,
d\Omega_+. \label{derplus}
\end{eqnarray}

Let us compute now the derivative w.r.t. $g_k$. We have:
\begin{eqnarray}
\frac{\partial y}{\partial g_k} &=&\frac{z^k W'}{y}+
\sum_{j=0}^{N-1}\frac{z^j}{2y}{\partial f_j \over \partial g_k}\nn\\
&=&z^k\left(1-\frac{f}{2 (W')^2}+\cdots\right)+
\sum_{j=0}^{N-1}\frac{z^j}{2y}{\partial f_j \over \partial g_k}\nn\\
&=&z^k\left(1- {f_{N-1}\over 2 g_N^2 z^{N+1} }  +
o(z^{-N-1})\right) +  \sum_{j=0}^{N-2}\frac{z^j}{2y}{\partial f_j
\over
\partial g_k} + \frac{z^{N-1}}{2y}  { \partial f_{N-1} \over
\partial g_k}. \label{derygk}
\end{eqnarray}
Suppose now that $0\leq k \leq N-1$. Then, in eq.\refs{derygk},
the first term is equal to $z^k + (\hbox{holomorphic terms})$, the
second term is holomorphic and the third term vanishes by use of
eq.\refs{derfn1}. Thus, we have ${\partial y \over \partial  g_k}
= z^{k} + (\hbox{holomorphic terms})$. If $k=N$, it is easy to see
that we come to the same conclusion thanks again to
eq.\refs{derfn1}. Furthermore,
\begin{equation}
\oint_{A_i}\frac{\partial y}{\partial g_k} =
-4\pi i\frac{\partial S_i}{\partial g_k}=0
\end{equation}
and thus
\begin{equation}
\frac{\partial y}{\partial g_k}=d\Omega_k. \label{dergk}
\end{equation}
We conclude that $y$ is homogeneous of degree one in
$(\alpha_i,\alpha_+,g_k)$ (and thus in $(S_i,g_k)$) by comparing
eq.\refs{newwhatisy} with the results \refs{derai}, \refs{derplus}
and \refs{dergk}.


\begin{thebibliography}{999}
\bibitem{VY1} G. Veneziano and S. Yankielowicz, Phys. Lett. B {\bf  113} (1982) 231.

\bibitem{1Inst} V.A. Novikov, M.A. Shifman, A.I. Vainshtein and
 V.I. Zakharov,  Nucl. Phys. B {\bf  260} (1985), 157
 (Yad. Fiz. 42, 1499); M.A. Shifman and A.I. Vainshtein,
Nucl. Phys. B {\bf  296}, (1988), 445 (Sov. Phys. JETP 66, 1100);
D. Amati, G.C. Rossi and G. Veneziano, Nucl. Phys. B {\bf  249}
(1985), 1; J. Fuchs and M.G. Schmidt, Z. Phys. {\bf C30} (1986),
161; D. Amati, K. Konishi, Y. Meurice, G.C. Rossi and G.
Veneziano, Phys. Rep. {\bf 162} (1988) 169.

\bibitem{Pouliot} D. Finnell and P. Pouliot, Nucl. Phys. B {\bf  453}
 (1995), 225 [arXiv:hep-th/9503115].

\bibitem{9407087}N. Seiberg and E. Witten, Nucl. Phys. B {\bf  426}
 (1994), 19 [arXiv:hep-th/9407087];
 Erratum-ibid, B {\bf  430} (1994) 485; N. Seiberg and
 E. Witten, Nucl. Phys. B {\bf  431} (1994), 484
 [arXiv:hep-th/9408099]; A. Klemm, W. Lerche, S. Theisen
 and S. Yankielowicz,
Phys. Lett. B {\bf  344} (1995) 169 [arXiv:hep-th/9411048]; A.
Klemm, W. Lerche and S.Theisen, Int. J. Mod. Phys. {\bf A11}
(1996), 1929 [arXiv:hep-th/9505150]; P. C. Argyres, A. E. Faraggi,
Phys. Rev. Lett. 74 (1995), 3931 [arXiv:hep-th/9411057].




\bibitem{West}P.S. Howe and P.C. West, Nucl. Phys. B {\bf 486} (1997),
  425 [arXiv:hep-th/9607239].

\bibitem{MSW} M. Magro, I. Sachs and S. Wolf, Annals Phys. {\bf 298}
  (2002), 123 [arXiv:hep-th/0110131].

\bibitem{9610026}
G.~Bonelli, M.~Matone and M.~Tonin,
Phys.\ Rev.\ D {\bf 55} (1997) 6466 [arXiv:hep-th/9610026]; R.
Flume, M. Magro, L.O'Raifeartaigh, I. Sachs and O. Schnetz, Nucl.
Phys. B {\bf  494} (1997), 331 [arXiv:hep-th/9611123].

\bibitem{FMRS2}M. Magro, L. O'Raifeartaigh and I. Sachs,
Nucl. Phys. B {\bf  508} (1997), 433 [arXiv:hep-th/9704027].

\bibitem{0103067}F. Cachazo, K. Intriligator and C. Vafa,
Nucl. Phys. B {\bf  603} (2001), 3 [arXiv:hep-th/0103067]

\bibitem{0103067b} F.~Cachazo and C.~Vafa, [arXiv:hep-th/0206017].

\bibitem{DV} R. Dijkgraaf and C. Vafa, [arXiv:hep-th/0208048].

\bibitem{0211170}F. Cachazo, M.R. Douglas, N. Seiberg and E. Witten,
JHEP {\bf 0212} (2002), 071 [arXiv:hep-th/0211170].

\bibitem{kacru} S. Kachru, A.~Klemm, W.~Lerche, P.~Mayr and C.~Vafa,
Nucl. Phys.  {\bf B459} (1996), 537 [arXiv:hep-th/9508155].


\bibitem{9609239} 
  S.~Katz, A.~Klemm and C.~Vafa,
  Nucl.\ Phys.\ B {\bf 497} (1997) 173
  [arXiv:hep-th/9609239].

\bibitem{9703166} A.~Klemm, W.~Lerche, P.~Mayr, C.~Vafa and
N.~P.~Warner, Nucl. Phys.  {\bf B477} (1996), 746
[arXiv:hep-th/9604034]; E.~Witten,
  Nucl.\ Phys.\ B {\bf 500} (1997) 3
  [arXiv:hep-th/9703166].


\bibitem{0206161} N. Nekrasov, Adv. Theor. Math. Phys. {\bf 7}
(2004) 831 [arXiv:hep-th/0206161]; N.~Nekrasov and S.~Shadchin,
  Commun.\ Math.\ Phys.\  {\bf 252} (2004) 359;
  [arXiv:hep-th/0404225]; ~W.~Moore, N.~Nekrasov and S.~Shatashvili,
  Commun.\ Math.\ Phys.\  {\bf 209} (2000) 77
  [arXiv:hep-th/9803265].

\bibitem{Flume} R. Flume, R. Poghossian and H. Storch, Mod. Phys. Lett.
{\bf A17} (2002), 327 [arXiv:hep-th/0112211]; R. Flume and R.
Poghossian, Int. J. Mod. Phys. {\bf A18} (2003), 2541
[arXiv:hep-th/0208176].

\bibitem{Bagger} J. Wess and J. Bagger,
{\em Supersymmetry and Supergravity}, Princeton University Press.

\bibitem{Konishi1} K. Konishi, Phys. Lett. B {\bf   135} (1984) 439.

\bibitem{Konishi2} K. Konishi and K. Shizuya, Nuovo. Cim. {\bf
A90} (1985), 111.

\bibitem{1001}
  S.~J.~Gates, M.~T.~Grisaru, M.~Rocek and W.~Siegel,
  Front.\ Phys.\  {\bf 58} (1983) 1
  [arXiv:hep-th/0108200].


\bibitem{0311238} P. Svrcek, JHEP {\bf 0410} (2004), 028
[hep-th/0311238].

\bibitem{svrcek2} P. Svrcek, JHEP {\bf 0408} (2004) 036
[hep-th/0308037].

\bibitem{0311181} K.~Intriligator, P.~Kraus, A.~V.~Ryzhov, M.~Shigemori and C.~Vafa,
  Nucl.\ Phys.\ B {\bf 682} (2004) 45
  [arXiv:hep-th/0311181].

\bibitem{Dorey} N. Dorey, V.V. Khoze and M.P. Mattis,
Phys. Rev. {\bf D54} (1996), 2921 [arXiv:hep-th/9603136];  N.
Dorey, T.J. Hollowood, V.V. Khoze and M.P. Mattis, Phys. Rept.
{\bf 371} (2002), 231 [arXiv:hep-th/0206063].

\bibitem{Yung} A. Yung, Nucl. Phys. {\bf B485} (1997), 38
[arXiv:hep-th/9605096].

\bibitem{9506102} M.~Matone,
  Phys.\ Lett.\ B {\bf 357} (1995) 342
  [arXiv:hep-th/9506102].

\bibitem{9805127} J.~D.~Edelstein, M.~Marino and J.~Mas,
  Nucl.\ Phys.\ B {\bf 541} (1999) 671
  [arXiv:hep-th/9805172].

\bibitem{0405117}  G.~Bertoldi, S.~Bolognesi, M.~Matone, L.~Mazzucato and Y.~Nakayama,
  JHEP {\bf 0405} (2004) 075
  [arXiv:hep-th/0405117].

\bibitem{0211017} R. Dijkgraaf, M.T. Grisaru, C.S. Lam,
C. Vafa and D. Zanon, Phys. Lett. B {\bf  573} (2003), 138
[arXiv:hep-th/0211017].

\bibitem{0206255} R. Dijkgraaf and C. Vafa, Nucl. Phys. B {\bf  644}
 (2002), 3 [arXiv:hep-th/0206255].

\bibitem{Russie} A.Gorsky, A.Marshakov, A.Mironov and A.Morozov,
Nucl.Phys. {\bf B527} (1998), 690 [arXiv:hep-th/9802007].

\bibitem{RBLR} P. Griffiths and J. Harris,
{\em Principles of Algebraic Geometry}, A. Wiley.

\bibitem{Naculich} S. Naculich, H. Schnitzer and N. Wyllard,
 Nucl.Phys. {\bf B651} (2003), 106 [arXiv:hep-th/0211123];
 M. Gomez-Reino, S.G. Naculich and H. J. Schnitzer,
 JHEP {\bf 0404} (2004), 033 [arXiv:hep-th/0403129].

\bibitem{9510129} J.~Sonnenschein, S.~Theisen and S.~Yankielowicz,
  Phys.\ Lett.\ B {\bf 367} (1996) 145
  [arXiv:hep-th/9510129].

\bibitem{9609104} K.~Ito and N.~Sasakura,
  Mod.\ Phys.\ Lett.\ A {\bf 12} (1997) 205
  [arXiv:hep-th/9609104].

\bibitem{Ito}
  K.~Ito and N.~Sasakura,
  Phys.\ Lett.\ B {\bf 382} (1996) 95
  [arXiv:hep-th/9602073]; E.~D'Hoker, I.~M.~Krichever and D.~H.~Phong,
  Nucl.\ Phys.\ B {\bf 489} (1997) 179
  [arXiv:hep-th/9609041].

\bibitem{FT} F. Fucito and G. Travaglini, Phys.Rev. {\bf D55}
(1997), 1099 [arXiv:hep-th/9605215].
\end{thebibliography}
\end{document}